# Enhanced and modulable induced superconducting gap and effective Landé *g*-factor in Pb-InSb hybrid devices

Guoan Li,[1,5,*] Xiaofan Shi,[1,5,*] Ziwei Dou,[1,*] Guang Yang,[1] Jiayu Shi,[1] Marco Rossi,[4] Ghada Badawy,[4] Yuxiao Song,[1,5] Ruixuan Zhang,[1,5] Yupeng Li,[1,9] Zhiyuan Zhang,[1,5] Anqi Wang,[1] Xingchen Guo,[1,5] Xiao Deng,[1,5] Bingbing Tong,[1,6] Peiling Li,[1,6] Zhaozheng Lyu,[1,6] Guangtong Liu,[1,6,8] Fanming Qu,[1,5,6,8] Erik P. A. M. Bakkers,[4] Michał P. Nowak,[2,†] Paweł Wójcik,[3,‡] Li Lu,[1,5,6,8,§] and Jie Shen[1,6,7,∥]

[1]Beijing National Laboratory for Condensed Matter Physics, Institute of Physics, Chinese Academy of Sciences, Beijing 100190, China
[2]AGH University of Krakow, Academic Centre for Materials and Nanotechnology, al. A. Mickiewicza 30, 30-059 Krakow, Poland
[3]AGH University of Krakow, Faculty of Physics and Applied Computer Science, al. A. Mickiewicza 30, 30-059 Krakow, Poland
[4]Department of Applied Physics, Eindhoven University of Technology, 5600 MB Eindhoven, The Netherlands
[5]School of Physical Sciences, University of Chinese Academy of Sciences, Beijing 100049, China
[6]Songshan Lake Materials Laboratory, Dongguan 523808, China
[7]Beijing Academy of Quantum Information Sciences, Beijing 100193, China
[8]Hefei National Laboratory, Hefei 230088, China
[9]Hangzhou Key Laboratory of Quantum Matter, School of Physics, Hangzhou Normal University, Hangzhou 311121, China

[*]These authors contributed equally to this work.
[†]Contact author: mpnowak@agh.edu.pl
[‡]Contact author: Pawel.Wojcik@fis.agh.edu.pl
[§]Contact author: lilu@iphy.ac.cn
[∥]Contact author: shenjie@iphy.ac.cn

**ABSTRACT**. The hybrid system of a conventional superconductor (SC) on a semiconductor (SM) nanowire with strong spin-orbit coupling (SOC) represents a promising platform for achieving topological superconductivity and Majorana zero modes (MZMs) towards topological quantum computation. While aluminum (Al)-based hybrid nanowire devices have been widely utilized, their limited superconducting gap and intrinsic weak SOC as well as small Landé $g$-factor may hinder future experimental advancements. In contrast, we demonstrate that lead (Pb)-based hybrid quantum devices exhibit a remarkably large and hard proximity-induced superconducting gap, exceeding that of Al by an order of magnitude. By exploiting electrostatic gating to modulate wavefunction distribution and SC–SM interfacial coupling, this gap can be continuously tuned from its maximum value (~1.4 meV, matching the bulk Pb gap) down to nearly zero while maintaining the hardness. Furthermore, magnetic-field-dependent measurements reveal a radial evolution of the gap structure with anti-crossing feature, indicative of strong SOC and huge effective $g$-factors up to 76. These findings underscore the superior functionality of Pb-based hybrid systems, significantly advancing their potential for realizing and stabilizing MZMs and the further scalable topological quantum architectures.

Majorana zero modes (MZMs), characterized by non-Abelian braiding statistics and topological protection against local perturbations [1,2], hold great promise for fault-tolerant topological quantum computation [3-5]. Theoretically, MZMs can emerge at the edge or vortex cores of topological superconductors [6-8]. Under favorable conditions in III-V semiconductor (SM) nanowires (such as InAs or InSb nanowires) with strong spin-orbit coupling (SOC) and large Landé $g$-factor, proximity-induced superconductivity from conventional superconductors (SC) combined with controlled chemical potential tuning and Zeeman field engineering induces a topological phase transition, resulting in the MZMs localized in the nanowire ends [9,10]. In addition, recently, the experimental progress of the artificial Kitaev chain for poor man's MZMs also based on the SC-SM hybrid nanowires has accelerated rapidly [11-14]. Driven by advanced material growth techniques, precise gate control capabilities and mature theoretical frameworks for braiding MZMs and designing topological qubits [5,15,16], these hybrid systems have become the most extensively studied platform for MZMs and topological quantum computation.

Among these, aluminum (Al) is the most widely adopted superconducting material in the present hybrid devices [17,18] due to the high-quality hetero-interfaces with oxide-free semiconductors and the resulting induced superconducting hard gap, which is signified by the large ratio between the above-gap and in-gap conductance (namely the hardness) in tunneling spectroscopy measurements [19-21] and the efficient 2e-periodic Coulomb-blockaded Cooper pair transport in hybrid island [22,23]. However, bulk Al suffers from intrinsically limited superconducting characteristics, such as the small superconducting energy gap ($\Delta$~0.2 meV) and low critical temperature ($T_C$~1.1 K) [24]. Recent non-local measurement shows the possible topological gaps in Al-InAs devices as low as 20-60 $\mu$eV [25]. Lead (Pb), in contrast, demonstrates superior superconducting properties with a critical temperature ($T_C$~7.2 K) and an intrinsic superconducting gap ($\Delta$~1.4 meV) approximately six-fold larger [24]. These characteristics provide enhanced robustness against thermal excitation, Zeeman-field and SOC strength fluctuations, as well as material disorder. Furthermore, due to Al's intrinsic weak SOC, the hybridized wavefunction suffers from a weak SOC and small $g$-factor in the strong SC-SM coupling limit, whereas a small proximity-induced superconducting gap in the weak-coupling limit [26,27]. Therefore, Al-based devices face a fundamental trade-off between achieving substantial induced superconducting gaps and preserving sufficient SOC/$g$-factor [28-30]—the two critical requirements for MZMs formation, while Pb-based devices simultaneously satisfies these conditions due to the intrinsic strong SOC and large $g$-factor. Experimental verification in magnetic atomic chain systems has confirmed Pb's ability to induce spin-orbit coupling via proximity effects [31,32].

Current advancements in Pb-InAs hybrid nanowire systems include direct measurements of robust superconducting gaps through tunneling spectroscopy and 2e-periodic Coulomb-blockaded Cooper pair transport in islands [33]. Gate-tunable supercurrent in Pb-PbTe Josephson junctions [27] has further demonstrated the potential for realization of hybrid superconducting devices with Pb superconductor. However, key research gaps exist: quantitative investigations into the tunable properties of proximity-induced gaps, SOC and effective $g$-factor, which are related to the overall controllability of the hybridized wavefunction depending on varying SC-SM coupling by tuning electrostatic gate, are still lacking despite theoretical expectations to solve the trade-off of Al's devices as mentioned above. These unresolved challenges represent critical future research direction for unlocking the full potential of Pb-based hybrid architectures in topological quantum technologies.

In this Letter, we fabricate Pb-InSb hybrid devices and perform the tunneling spectroscopy measurements that reveal three critical characteristics. First, the tunnel differential conductance exhibits a proximity-induced superconducting gap with a hardness reaching two order of magnitudes and an amplitude exceeding 1 meV. Second, the precise control of electrostatic gate voltages enables continuous tuning of the gap magnitude—from its peak value (~1.4 meV, matching bulk Pb) to close to zero—without compromising its hardness. This remarkable tunability arises from gate-induced modulation of wavefunction distributions across the nanowire cross-section and the corresponding variations in SC-SM interface coupling. Third, magnetic-field-dependent measurements of the superconducting gap reveal a closure-reopening sequence accompanied by radial-line structure. Comparison with theoretical models elucidates that the observed radial gap evolution originates from the coexistence of the SOC and distinct g-factors from multiple modes interacting with Pb. This enhances the effective g-factor to the values much higher that of Al-based devices, even exceeding that of intrinsic bulk InSb. These findings establish Pb-InSb hybrid devices as a versatile platform for engineering tunable superconducting gaps and amplified g-factors, thereby greatly advancing prospects for realizing MZMs in scalable topological quantum systems.

Figures 1(a) and 1(d) show the scanning electron microscope (SEM) images of two typical Pb-InSb hybrid devices, labeled Device 1 (D1) and Device 2 (D2), respectively. The fabrication details are listed in Appendix A. Transport measurements for both devices are performed in the dilution refrigerator with a base temperature of approximately 10 mK. We employ the standard DC+AC lock-in technique to obtain the tunnel spectroscopy, which directly reflects the quasiparticle density of states in the proximitized section of the nanowire. Figure 1(b) shows the differential conductance of D1 as a function of bias voltage $V_{bias}$ and back-gate voltage $V_{BG}$. The red vertical line cut shown in Fig. 1(c) at the selected gate voltage in logarithmic scale shows the ratio of above-gap and subgap conductance as high as two orders of magnitude, confirming the hardness of the induced superconducting gap. The remarkable in-gap conductance suppression is comparable to the hard gap in Al-based hybrid devices [19-21]. A similar hardness of the induced gap is also demonstrated in D2—an island in which we can eliminate the Coulomb blockade by opening the nonlocal barrier to perform tunnel measurement on the local barrier [34] and measure an even larger gap up to ~1.4 meV (red cut in Fig. 1(f)). The measurements of the conductance curves are swept from negative to positive bias, whereby the stronger self-heating effect [35,36] at the negative bias causes an asymmetry of the tunneling spectrum with respect to the zero bias (see Supplemental Material [37]). To accurately determine the size of the superconducting gap, it is preferable to extract the coherent peak at positive bias. This large induced gap, in particular in D2, which is six-fold larger than that of Al, will greatly expand the topological gap whose upper limit is determined by the induced gap [25].

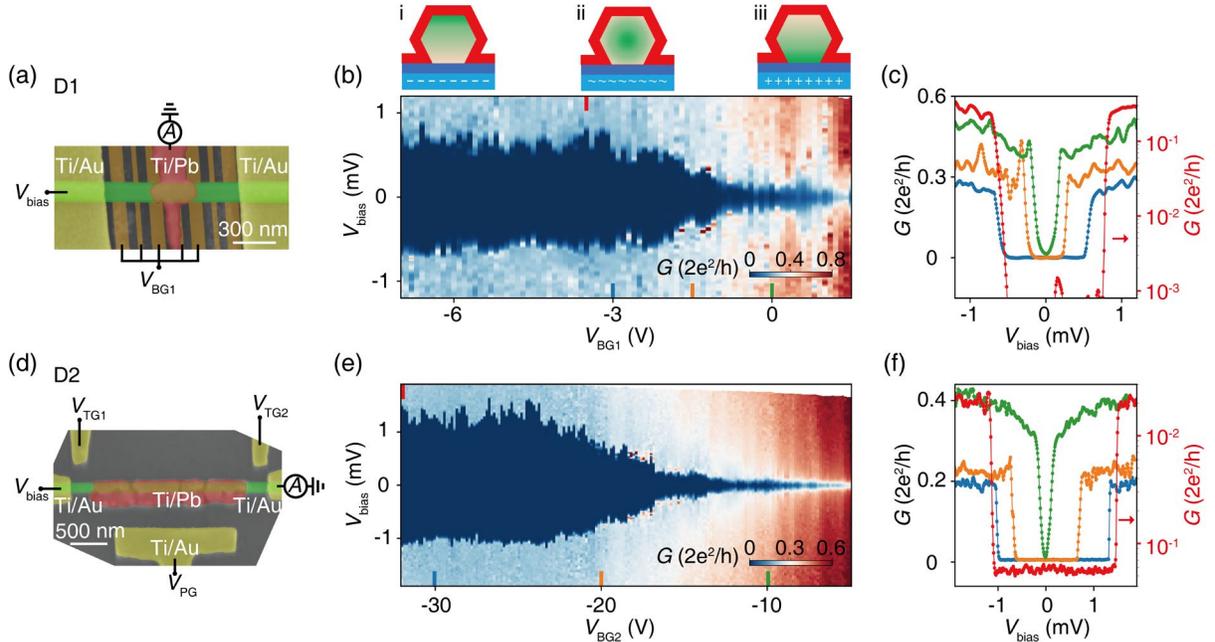

Fig. 1 Transport measurement of InSb-Pb normal metal-superconductor tunnel junctions of two devices D1 and D2. (a, d) False-color SEM image of D1 and D2. Ti/Pb electrodes (red) and Ti/Au electrodes (yellow) contact the InSb nanowire (green), forming normal metal-superconductor tunnel junctions. Although both devices have a continuous Pb film, due to thermal contact with the substrate holder (~130 K) cooled via liquid nitrogen, the Pb film on D1 (~110 nm) is more uniform in morphology than the room temperature grown Pb film on D2 (~28 nm). (b, e) Conductance as a function of $V_{bias}$ and $V_{BG}$ for the two devices. The top panels show the schematics of the three coupling regimes. The substrate acting as the back gate is indicated by blue, while the superconducting region and the spatial charge distribution in the nanowire are indicated by red and green, respectively. (c, f) The color (red) curves show conductance as a function of $V_{bias}$ with linear (log) Y axis, corresponding to the vertical linecuts at different $V_{BG}$ values in (b,e) as indicated by the color(red) bars. The full linecuts are listed in Supplemental Material [37].

Moreover, we find the induced superconducting gap can be gate-tunable to a large extend, as shown in Figs. 1(b) and 1(e). The vertical line cuts at different $V_{BG}$ in Figs. 1(c) and 1(f) also reveal the variation of the induced gap, ranging from approximately 0 to 0.6 meV for D1 and from 0 to 1.4 meV for D2, respectively (see the full waterfall data in Supplemental Material [37]). The large gap always shows up at relatively negative gate value and shrinks at more positive gate value. This reveals that the electrostatic gate not only controls the chemical potential but also effectively modulates the electric field and thus the wavefunction distribution in the cross-section of the hybrid structure, influencing the coupling between nanowires and superconductors and the final hybridized wavefunction. As shown in previous Al-based devices and theoretical calculation [26,38-40], when the back-gate voltage goes to negative value, more electrons are pushed by the electric field towards the interface and close to the bulk Pb, leading to stronger coupling and a larger induced superconducting gap as illustrated by the top left panel (i) in Fig. 1(b). Conversely, positive gate voltage attracts more electrons towards the facet underneath and far away from Pb, resulting in weaker coupling and smaller gap as shown by the top right panel (iii) in Fig. 1(b). This tunability has been observed in Al-hybrid devices, but never reaches such a large variation [26]. In addition, the difference in the maximum of the induced gap in the two devices may originate from the different band bending strength at the interface, which induce different coupling strength and tunability of the electrostatic gate [39-42]. Interestingly, with such a great tunability, the gap is always hard with negligible in-gap states, totally different from the Al case where the smaller gap at positive gate always becomes soft [26] (see Figs. 1(c, f) and Supplemental Material [37]). We can exclude the gap here as the Coulomb gap for the reasons given in Appendix B. This large hard induced gap combined with the highly effective electric field response greatly extends the experimental parameter space, making the Pb-InSb hybrid device a promising platform to search for MZMs, in particular if it has the strong SOC and $g$-factor as expected.

Then to study the SOC and $g$-factor, we apply a magnetic field perpendicular to the substrate for the two devices. In Fig. 2, we show the magnetic-field-dependent tunneling spectroscopy at some typical gate voltages, accompanied with different induced gap sizes corresponding to different SC-SM coupling strengths and hybridization. Clearly, multiple radial resonance lines at finite bias with different slopes can be observed in the conductance map as pointed out by the colored dashed lines at all gate values. We define the effective $g$-factor by the slope of these states as a function of the magnetic field and show the results in the legend. (It should be noted that the lines at low magnetic field sometimes bend as shown by black dashed lines in Fig. 2(d) but has unified slopes at high magnetic field, so we use the latter to extract $g$-factor. This is reasonable according to the calculation discussed later in Fig.4(b).) The slopes of these radial lines are significantly different, corresponding to the extraction of different effective $g$-factors (~7 to 76, listed in the bottom-left corner of each figures in Fig. 2, also included in Fig. 3(e)). Some of these $g$-factors are much larger than the Landé $g$-factor of bulk InSb (40 to 50) [43]. Intriguingly, a large $g$-factor is still kept in the large induced gap in the strong-coupling limit. This contrasts sharply with the Al's hybrid device, in which strongly coupling to Al leads to a smaller $g$-factor due to Al's negligible $g$-factor (~2) [44]. Such a large effective $g$-factor is exactly desirable in nanowire hybrid systems for the lower-field trivial-to-nontrivial topological phase transition, whose closure is limited by the critical magnetic field of parent superconductor, and thus enlarges the entire topological regime. It is also interesting that the lines tangent to the resonance do not originate at the gap edge in zero-field as pointed out by green arrow in Fig. 2(b). In some of the maps, we observe the re-openings of gap after its closure as well as anti-crossings structure, which is different from the BCS gap [52] and probably indicative of the SOC. We also perform magnetic-field-dependent measurement from three different directions (see Supplemental Material [37]), which reveals the gap evolution is almost isotropic, and thus exclude the anisotropic orbital effect here.

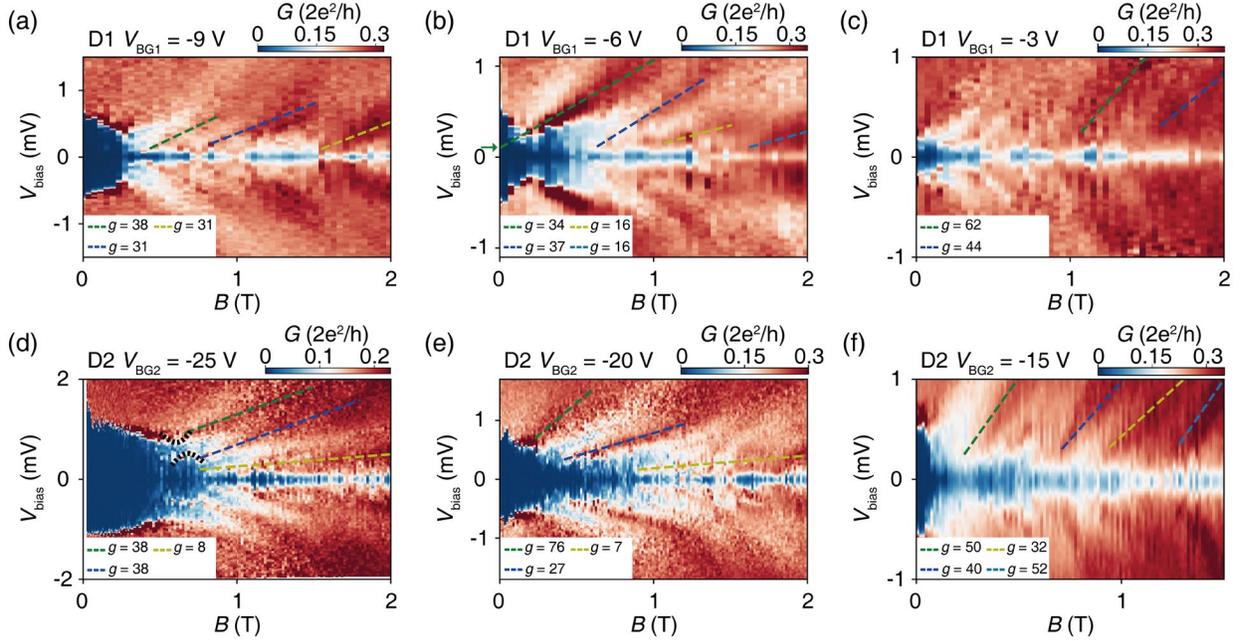

Fig. 2 Transport data of Pb-InSb devices in an out-of-plane magnetic field. (a-c) Conductance as a function of $V_{bias}$ and magnetic field $B$ for D1. The dashed lines show the evolution of conduction as a function of the magnetic field, and the corresponding effective g-factors are shown in the white box. (d-f) The similar data for D2.

To compare with the experimental results, we conduct tunnel spectroscopy calculations for the hybrid device with a tunnel barrier in the presence of an out-of-plane magnetic field. We consider a one-dimensional nanowire described by the Hamiltonian,

$$H = \left(\frac{\hbar^2 k_x^2}{2m^*} - \mu - V(x)\right)\sigma_0\tau_z + \Delta(x)\sigma_0\tau_x - \alpha k_x\sigma_y\tau_z + E_z\sigma_z\tau_0 \quad (1)$$

The system consists of a proximitized nanowire (with $\Delta$ - the induced superconducting gap) connected to a normal electrode through a potential barrier (described by the $V(x)$ potential) and a superconducting drain electrode. The details of the numerical method are given in Supplemental Material [37]. A Rashba-type spin-orbit interaction, with strength governed by the parameter $\alpha$, is incorporated within the nanowire, along with a Zeeman interaction characterized by the energy $E_z = \frac{1}{2}g\mu_B B$. We begin by examining the case without spin-orbit coupling. Figure 3(a) presents the conductance map obtained when the potential barrier is tuned to the tunneling regime. Notably, conductance peaks can be observed, precisely located at the gap edge when the magnetic field is zero. As the Zeeman interaction separates the spin-opposite bands in energy, the corresponding peaks in the conductance map exhibit splitting. Upon meeting of the inner peaks, the superconducting gap closes. Similar results are observed in some Al-based hybrid nanowire experiments [45]. As expected, the slope of the induced gap evolution is exactly the g-factor assumed in the Hamiltonian.

Further, Pb, a heavy element superconductor that benefits from strong intrinsic spin-orbit coupling, is verified to possess strong Rashba spin-orbit coupling in the monolayer [46,47]. We therefore proceed to consider the case of non-zero $\alpha$ to explore the potential characteristics of Pb-based hybrid devices. We observe that the slope of the inner resonance line now deviates from that obtained without SOC, resulting in a modified |g|-factor, reduced from 50 to 46 (Fig. 3(b)). Another important feature is that the tangent line fitted to the resonance line does not precisely originate at the position of coherent peaks in the zero field, which is consistent with the experiment. To better understand the behavior of the resonance lines, we have derived the analytical formula in Appendix C, which shows clearly the anti-crossings between the opposite spin bands resulting in the gap re-openings and anti-crossings observed in experiments.

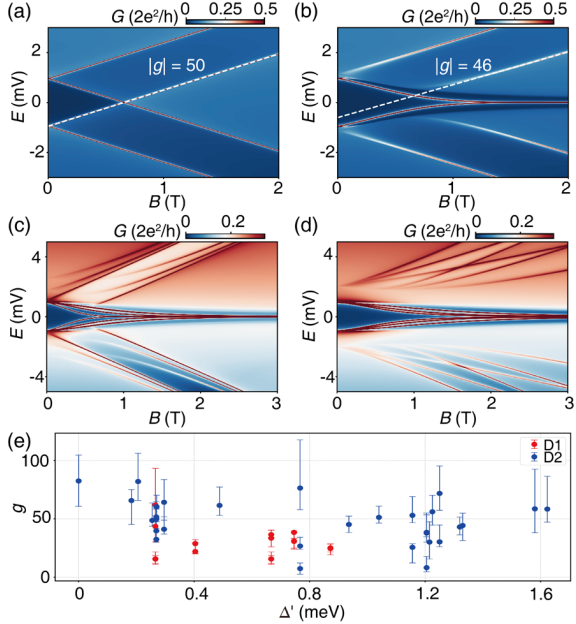

Fig. 3 Calculated tunnel spectroscopy of the hybrid device in an out-of-plane magnetic field. (a) Tunnel conductance map without SOC. (b) Tunnel conductance map considering SOC with $\alpha = 10$ meV nm. (c) Conductance obtained for the same g-factor ($g = -80$) of the modes but different $\alpha$ ($\alpha_1 = 10$ meV nm, $\alpha_2 = 20$ meV nm, $\alpha_3 = 45$ meV nm). (d) Conductance obtained for the same $\alpha$ ($\alpha = 30$ meV nm) of the modes but different g-factors ($g_1 = -40$, $g_2 = -60$, $g_3 = -80$). (e) Statistical graph of effective g-factor and induced gap with error bars obtained in the experiment.

We now consider a case where the wire is populated by three modes with each having different SOC constant and g-factor. The model bases on block diagonal Hamiltonian with diagonal terms Eq. (1) and off-diagonal mixing terms—see Supplemental Material [37] for details. Figure 3(c) illustrates the case where three modes with the same g-factor but different $\alpha$ values are considered. We see that at low magnetic fields the spin-orbit coupling leads to different renormalization of the resonance line slope from which three different g-factors can be extracted. However, at large magnetic fields, all three curves have the same slope. On the other hand, in Fig. 3(d), we keep $\alpha$ the same for all modes but vary the g-factor. Similarly, spin-orbit coupling can lead to modifications of the g-factor. In contrast, the resonance lines do not exhibit a common slope at large magnetic fields, consistent with experimental observations, suggesting that the modes in the nanowire have different (bare) g-factors.

Last, we compare and analyze the renormalization of the effective g-factors. As discussed above, a wide range of g-factors may be attributed to g-factor renormalization brought about by coupling to Pb. In order to compare the hybridization strengths at each gate voltage, the sizes of the induced superconducting gap at zero magnetic field are used to quantify them. We then statistically record the g-factor from the evolution data of both devices with magnetic field at various back-gate voltages. The extraction of the g-factor and the method of obtaining the corresponding error bars are discussed in detail in the Supplemental Material [37]. Figure 3(e) displays the final statistical diagram. From the case of the maximum g-factor at each gate voltage, there is a tendency for the g-factor values to increase with enhanced nanowire and superconductor coupling. Also, with a smaller induced gap maximum, the g-factors of D1 are overall lower than that of D2. Theoretically, confinement effects in the nanowire structure would result in a reduced g-factor of the nanowire concerning the bulk [48,49], but spin-orbit coupling can lead to the enhancement of g-factors [50]. We believe that the persistence of this large g-factor, even in the strong coupling region, which is significantly different from Al's case, may be attributed to the non-negligible Rashba spin-orbit coupling of Pb. More discussion about the comparison between theory and experiment is concluded in Appendix C. Apparently, this innovative discovery reconciles the longstanding trade-off in Al-based devices between significant proximity-induced superconducting gaps and strong spin-orbit coupling/g-factor.

In summary, we report the observation of a large and hard induced superconducting gap in Pb-InSb hybrid devices. The superconducting gap's magnitude—orders of magnitude larger ($\sim 1$ meV) than aluminum-based systems ($< 200$ μeV)—will greatly extend experimental parameter regimes available for MZMs studies. Moreover, it also presents a wide range of variation through gate-voltage modulation, evidencing remarkable adaptability in controlling electron wavefunction distribution at the cross-section and the SC-SM hybridization. Last but not least, this platform exhibits a non-negligible SOC and enhanced $g$-factor at the entire coupling regime, which are inherited from the Pb superconductor and crucial ingredients for topological quantum state engineering. Collectively, these results establish Pb as a superior parent superconductor compared to Al, offering improved resilience to thermal fluctuations, magnetic interference and disorder interruption. This advancement provides a compelling experimental platform for realizing topological quantum systems, motivating further investigations into optimizing Pb-based hybrid devices for the scalable implementation of Majorana-based qubit architectures.

## ACKNOWLEDGMENTS

The work of Z.D. and J.S. was supported by the Young Scientists Fund of the National Natural Science Foundation of China (Grant No. 2024YFA1613200) The work of J.S., L.L., F.Q. and G.L. were supported by the National Key Research and Development Program of China (Grant Nos. 2023YFA1607400), the Beijing Natural Science Foundation (Grant No. JQ23022), the Strategic Priority Research Program B of Chinese Academy of Sciences (Grant No. XDB33000000), the National Natural Science Foundation of China (Grant Nos. 92365302, 12174430), and the Synergetic Extreme Condition User Facility (SECUF, https://cstr.cn/31123.02.SECUF). Y.L. acknowledges support from National Natural Science Foundation of China, Grant No. 12404154. The work of other authors were supported by the National Key Research and Development Program of China (Grant Nos. 2019YFA0308000, 2022YFA1403800, 2023YFA1406500, and 2024YFA1408400), the National Natural Science Foundation of China (Grant Nos. 12274436, 12274459), the Beijing Natural Science Foundation (Grant No. Z200005), and the Synergetic Extreme Condition User Facility (SECUF, https://cstr.cn/31123.02.SECUF). The work is also funded by Chinese Academy of Sciences President's International Fellowship Initiative (Grant No. 2024PG0003). M.P.N. and P.W. acknowledge support by the program "Excellence Initiative - research university" for AGH University of Krakow.

### Author Contributions

G.A.L., X.F.S., and Z.W.D. contributed equally to this work. J.S. conceived and designed the experiment. J.Y.S., X.F.S. and, G.A.L. fabricated devices with the help of G.Y. and Z.Y.Z., G.A.L., X.F.S. and J.Y.S. performed the transport measurements, supervised by G.Y., B.B.T., P.L.L., Z.Z.L., G.T.L, F.M.Q., Z.W.D., L.L. and J.S., G.A.L., G.Y., X.F.S., M.P.N., P.W. and J.S. analyzed the data. M.R., G.B. and E.P.A.M.B. provided InSb nanowires. M.P.N. and P.W. conducted tunnel spectroscopy calculations. G.A.L., Z.W.D., M.P.N., P.W. and J.S. wrote the manuscript, and all authors contributed to the discussion of results and improvement of the manuscript.

## APPENDIX A: FABRICATION DETAILS

With the aid of a micro-manipulator tip under the optical microscope, high mobility nanowires grown by metal organic vapor phase epitaxy (MOVPE) [51] are first transferred from the growth chip to the substrates covered by boron nitride (~20 nm) or silicon dioxide (~300 nm) as dielectric layers. And the titanium and gold (Ti/Au) electrodes, or heavily doped silicon underneath, function as back-gate electrodes. In order to obtain good contact and appropriate band bending at the hetero-interface, we use gentle argon milling to remove the oxide layer of the nanowire and induce an smooth and accumulation layer on the surface before depositing the films, as proved by our previous devices [41]. The uncovered section of the nanowire between the Ti/Au (yellow) normal metal electrode and the Ti/Pb (red) superconductor electrode serves as a tunnel barrier, which is tunable by the electrostatic gate. The reason that we add a Ti wetting layer after argon milling and before Pb film deposition is to avoid the large lattice mismatch between Pb and InSb [52-54]. Besides, immediately after depositing the Pb shells, the aluminum oxide ($AlO_x$) is evaporated to prevent the oxidation and dewetting of the Pb. We notice D1 is a typical N-S-N device with barriers in-between to perform tunneling spectroscopy on each terminal while D2 is an island device with two barriers, for which we open the nonlocal barrier by applying the very positive value for TG2 and enable the tunneling spectroscopy via controlling local barrier by tuning TG1.

## APPENDIX B: COULOMB GAP EXCLUSION

While the tunable gap may appear to be an unintentional Coulomb-blockaded gap instead of the induced superconducting gap, it can be ruled out for the following four reasons: First, the closure of the gap depends on a wide range of gate voltage values, which is distinct from the Coulomb gap associated with much smaller gate voltage variation determined by charging energy and lever arm; Second, the maximum gap of D2 is nearly identical to the gap of the parent Pb, suggesting that it is more like proximity-induced superconducting gap; Third, the gap also saturates for a large range of gate variation for both devices, which is totally different from the diamond shape of Coulomb gap; Fourth, there are no observable excited states above the gap, which typically present in Coulomb-blockaded gaps.

## APPENDIX C: THEORETICAL CALCULATION

To better understand the behavior of the resonance lines, we analytically solve the translation-invariant system (with $k_x$ a good quantum number) described by the Hamiltonian in Eq. (1). We find that the positive energy solutions are given by:

$$E^{\pm} = \sqrt{\Delta^2 + E_z^2 \pm 2\Delta E_z + \frac{2\alpha^2 m^* \mu}{\hbar^2} - \frac{\alpha^4 m^{*2}}{\hbar^4}} \quad (2)$$

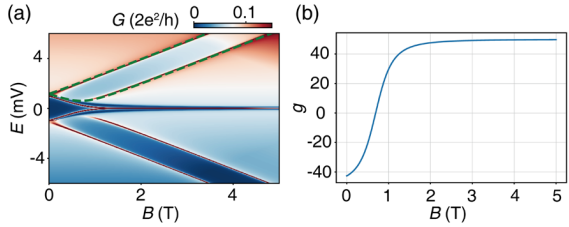

Fig. 4 Other calculations of tunnel spectroscopy. (a) Comparison of numerical results with analytical model (green dashed lines). (b) Extracted g-factor at different magnetic field.

Figure 4(a) presents a comparison between the numerical and analytical results (indicated by green dashed lines). At experimentally relevant small fields, a pronounced bending of the resonance line is observed. For sufficiently large $B$, the Zeeman terms $E_Z$ dominates over the constant terms introduced by the SOC in Eq. (2) and the renormalization effect becomes negligible. Figure 4(b) shows g-factors extracted from the analytical formula using $g = \left(\frac{\partial E}{\partial B}\right) \cdot \left(\frac{2}{\mu_B}\right)$ for the energy levels obtained from Eq. (2). In the conductance map, the increased conductance line starts to appear at around 1 T and in the g-factor plot we see a strong renormalization resulting from bending of the resonance line. As $B$ increases, $g$ approaches a saturation value, stabilizing around 5 T. Therefore, in the dispersion relation, there are anti-crossings between the opposite spin bands [55,56], resulting in the gap re-openings and anti-crossings observed in experiments.

To conclude, the experiment measures faint resonance lines with different slopes in the magnetic field and with tangents that do not originate at the gap edge at almost all gate values with different SC-SM coupling and hybridization. In the absence of spin-orbit coupling, the slope of these resonance lines is directly determined by the g-factor of the modes in the wire. We show that at low magnetic fields the spin-orbit interaction leads to a modification of the size of the gap in the dispersion relation in the nanowire. This in consequence leads to bending of the resonance lines from which modifies the extracted g-factor. The amount of deviation from the bare g-factor is determined mostly by the strength of the SOC (which is controlled both by $\alpha$ but also by the inverse wave vector of a given band at given energy), and the renormalization always makes the estimated g-factor smaller (at small fields). Moreover, a tangent line fitted to those bent resonances does not start at the gap edge at zero magnetic field, which is not the case without the SOC. Note that such lines were not observed in the wires proximitized by Al with negligible SOC. This means the distinguishable SOC, as well as large g-factor, appears in all the gate regime with different superconducting gap and SC-SM coupling, in particular for the strong coupling case when Al-based devices have negligible SOC and g-factor in spite with the relatively large gap.

# Supplemental Material Enhanced and modulable induced superconducting gap and effective Landé *g*-factor in Pb-InSb hybrid devices

**Analysis of self-heating effect, waterfall plot and magnetic-field-dependent measurement**

Measuring a large induced superconducting gap inevitably requires a substantial current, which introduces Joule heating and the possible subsequent in electron temperature [1,2]. Due to the variation in heating power under different bias currents, the sweeping method will affect the gap size. To quantify this self-heating effect, a typical conductance curve is chosen for analysis [Fig. S1(a)]. Figure S1(b) illustrates the *I-V* curve obtained by integrating the differential conductance. The Joule heating can then be calculated based on the current and voltage at each point [Fig. S1(c)]. The thermal power at high $V_{bias}$ is significantly higher than at low $V_{bias}$. Moreover, the thermal power above the order of pW will increase temperature above the dilution refrigerator's base temperature of 10 mK [2]. Therefore, when the tunnelling spectrum is swept from negative to positive bias, a smaller superconducting gap at the negative $V_{bias}$ due to the higher electron temperature can be expected. This phenomenon is consistent with the features of the tunnelling spectrum demonstrated in the main text. The full linecuts of Figs. 1(b) and 1(e) are shown in Fig. S2. In addition, the magnetic-field-dependent measurement data at different directions used to exclude the orbital effect are presented in Fig. S3.

**Single-mode effective model**

As the device D1 has a grounded superconductor that serves as a drain and D2 has a superconducting island with length much larger than the superconducting coherence length (~ 2 μm vs ~ 330 nm) we treat the experimentally studied devices as normal-superconductor junctions. We introduce one-dimensional model of the junction described by the Hamiltonian,

$$H = \left( \frac{\hbar^2 k_x^2}{2m^*} - \mu - V(x) \right) \sigma_0 \tau_z + \Delta(x) \sigma_0 \tau_x - \alpha k_x \sigma_y \tau_z + E_z \sigma_z \tau_0 \tag{S1}$$

The modeled system consists of a normal lead, a tunneling barrier (described by $V(x)$ potential) and a superconducting contact (with non-zero $\Delta(x)$). We include a spin-orbit interaction term (whose effective strength is controlled by $\alpha$ and which can originate from intrinsic nanowire spin-orbit coupling of the nanowire or from the nearby Pb shell) and a Zeeman interaction characterized by the energy $E_Z = \frac{1}{2} g \mu_B B$ for the magnetic field oriented out-of-plane. We discretize the Hamiltonian Eq. (S1) on a finite mesh with lattice spacing $a = 6$ nm and take the InSb effective mass $m^* = 0.014$ m (with *m* the free electron mass), chemical potential $\mu = 10$ meV and the induced gap $\Delta = 1$ meV. We consider one-site wide and 0.2 μeV high potential barrier to tune the system into tunneling regime. We calculate the conductance of the system tracing the electron ($R_{ee}$) and holes ($R_{he}$) reflected from the scattering region upon injections of the electrons from the normal electrode, following the formula

$$G(E) = \frac{e^2}{h} (N - R_{ee}(E) + R_{he}(E)) \tag{S2}$$

where the reflection coefficients $R_{ee}$, $R_{he}$ are obtained from the system scattering matrix obtained in the Kwant package [3] and *N* is the number of modes.

An example of the tunneling conductance obtained in the model is shown in Fig. S4(a) along with the dispersion relation of the superconducting lead (b). The results are obtained for $\alpha = 10$ meV nm. The gap edges seen in the spectroscopy map are renormalized by the spin-orbit interaction and corresponds to the energy gap present in the dispersion relation. We observe that the inner resonance line slope and the *g*-factor deviate from the one obtained in the absence of spin-orbit coupling and the extracted *g*-factor is modified from the assumed here |g| = 50 to |g| = 46. Moreover, the tangent line that we fit to the resonance line does not originate at zero energy.

**Multi-mode effective model**

To effectively describe the junction populated by more than one mode (here we assume three) we adopt a model described by the Hamiltonian

$$H_{\text{multimode}} = \begin{pmatrix} H_1 & \gamma\sigma_0\tau_z & \gamma\sigma_0\tau_z \\ \gamma\sigma_0\tau_z & H_2 & \gamma\sigma_0\tau_z \\ \gamma\sigma_0\tau_z & \gamma\sigma_0 & H_3 \end{pmatrix} \tag{S3}$$

$H_{1,2,3}$ are original single-mode nanowire Hamiltonians, Eq. (S1). In each of the $H_i$ Hamiltonian we consider unique values of the $\alpha_i$, $g_i$ constants. The off-diagonal $\gamma$ terms provide the mode mixing.

**Extraction of the *g*-factor**

Despite the evident linear traces of conductance in the 2D maps, the noise fluctuations obscure the exact positions of the corresponding conductance peaks. To enhance the extraction accuracy of the effective *g*-factor, we employ the function savgol.filter from the Python scipy.signal library to smooth the conductance curves [4]. Figure S5(a) shows that a typical conductance curve is successfully smoothed to remove the noise fluctuations while still retaining the important features of the original data. Subsequently, the function peak.finding from the scipy.signal library is used to identify the smoothed peaks [4]. Each conductance curve under different magnetic fields undergoes the above process, with the results depicted in the 2D map [Fig. S5(b)]. Due to the particle-hole symmetry in superconductors, we overlap the peaks identified at both the positive and negative bias voltage, and the aggregation of the peaks validates our hypothesis and confirm the accuracy of the extracted *g*-factor. Finally, we visually evaluate the values of the effective *g*-factor along with the error bars by the dashed lines in Fig. S5(b). Similar *g*-factor extractions are performed for data at other coupling strengths. The relevant statistics are discussed in the main text.

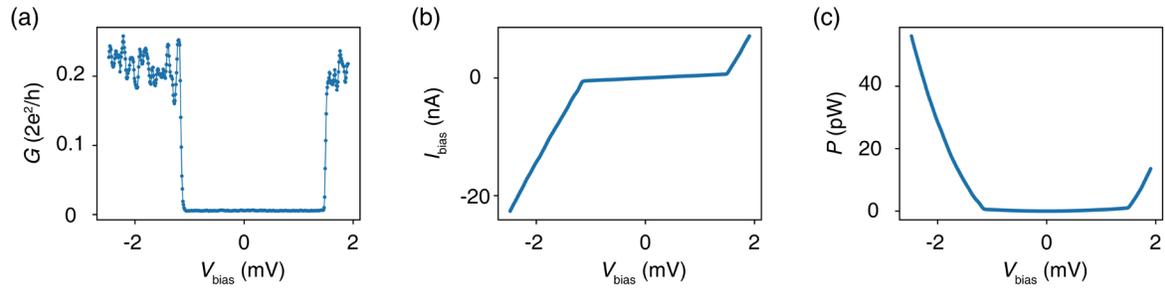

Fig. S1 Self-heating effect analysis. (a) A typical tunnelling spectrum showing the asymmetric superconducting gap. (b) $I$-$V$ curve after integration from (a). (c) Calculated Joule power as a function of $V_{bias}$. The Joule power at high $V_{bias}$ is significantly higher than at low $V_{bias}$.

(a)

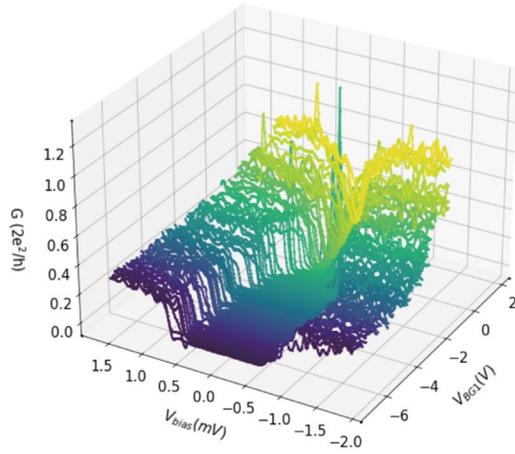

(b)

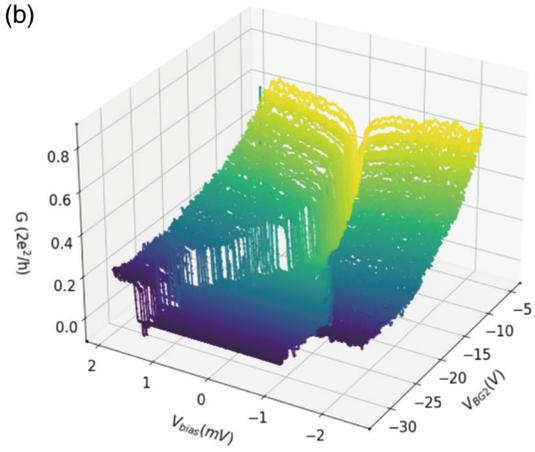

Fig. S2 Three dimension-waterfall plot of the conductance map. (a,b) Correspond to Fig. 1(b) and 1(e) of the main text, respectively.

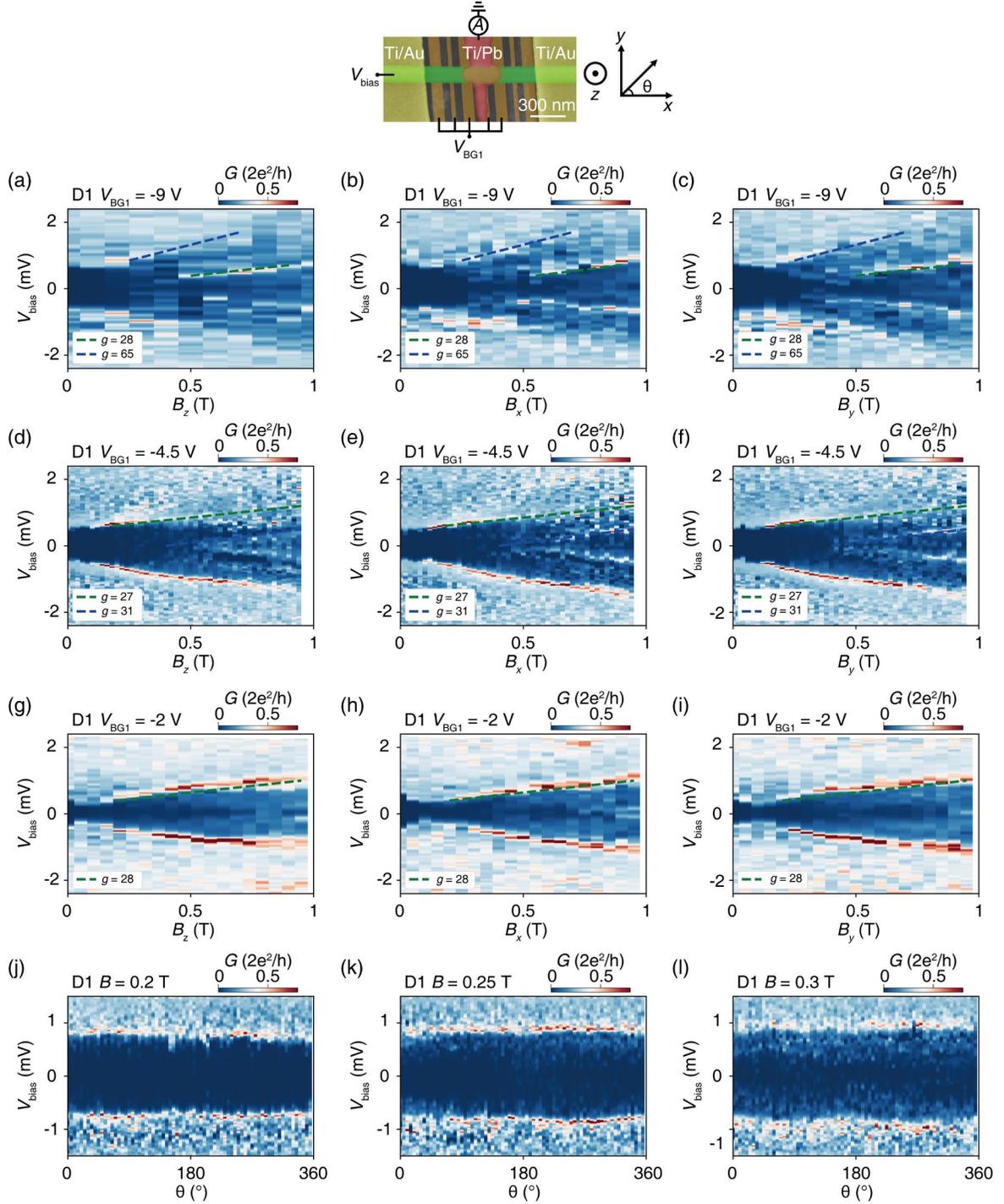

Fig. S3 Magnetic-field-dependent measurement at different directions. (a-c) Tunneling conductance as a function of $B_z$, $B_x$, and $B_y$ at $V_{BG1}$ = -9 V, respectively. (d-i) The similar data at $V_{BG1}$ = -4.5V, $V_{BG1}$ = -2 V. (j-l) Tunneling conductance as a function of the in-plane magnetic field angle at fixed magnitudes of 0.2 T, 0.25 T, and 0.3 T. Since the vector magnet of the dilution refrigerator has maximum applicable fields of 9 T, 1 T, and 1 T in the three directions respectively, comparisons are made with all three directions set to 1 T. However, this range essentially demonstrates negligible anisotropy.

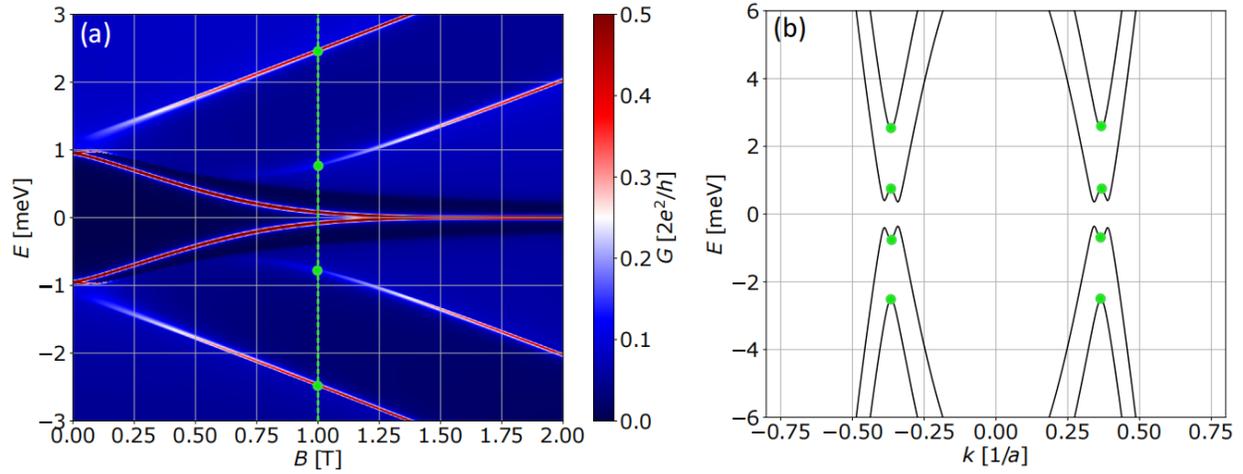

Fig. S4 Tunneling conductance obtained in the single-mode effective model. (a) Tunneling conductance map versus energy and the magnetic field with spin-orbit coupling included, $\alpha = 10$ meV nm and assumed $g = -50$. (b) Dispersion relation of the superconducting nanowire at $B = 1$ T. Green dots show the gap edges at which the conductance peaks appear in the map of panel (a).

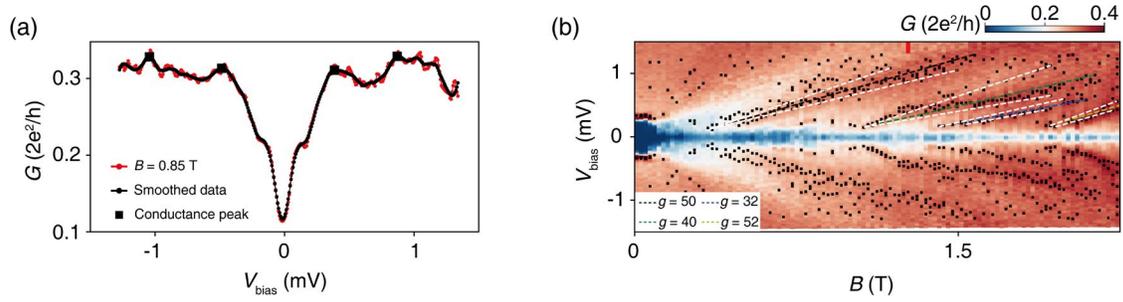

Fig. S5 Extraction of the *g*-factor. (a) Conductance as a function of $V_{bias}$ at $B = 0.85$ T for Device 2 (in red) and smoothed data (in black). Black square symbols mark the positions of the identified conductance peaks. (b) Conductance as a function of $V_{bias}$ and magnetic field $B$. The black square symbols also represent the conductance peaks. The dashed lines show the extracted effective *g*-factors (in color) and the error bars (in white).